# Brain Performance Analysis based on an Electroencephalogram Headset

Iuliana MARIN
Faculty of Engineering in Foreign Languages
University POLITEHNICA of Bucharest
Bucharest, Romania
marin.iulliana25@gmail.com

Ioana-Andreea DINESCU
Faculty of Engineering in Foreign Languages
University POLITEHNICA of Bucharest
Bucharest, Romania
ia.dinescu0@gmail.com

Teodora-Coralia DELEANU
Faculty of Engineering in Foreign Languages
University POLITEHNICA of Bucharest
Bucharest, Romania
teo.coralia97@yahoo.com

Lujain ALSHIKH SULAIMAN
Faculty of Engineering in Foreign Languages
University POLITEHNICA of Bucharest
Bucharest, Romania
loujain1997@gmail.com

Sarmad Monadel Sabree AL-GAYAR
Doctoral School of Automatic Control and Computers
University POLITEHNICA of Bucharest
Bucharest, Romania
sarm.moon85@gmail.com

Nicolae GOGA
Faculty of Engineering in Foreign Languages
University POLITEHNICA of Bucharest
Bucharest, Romania
n.goga@rug.nl

*Abstract*—Deficit of attention, anxiety, sleep disorders are some of the problems which affect many persons. As these issues can evolve into severe conditions, more factors should be taken into consideration. The paper proposes a conception which aims to help students to enhance their brain performance. An electrocephalogram headset is used to trigger the brainwaves, along with a web application which manages the input data which comes from the headset and from the user. Factors like current activity, mood, focus, stress, relaxation, engagement, excitement and interest are provided in numerical format through the use of the headset. The users offer information about their activities related to relaxation, listening to music, watching a movie, and studying. Based on the analysis, it was found that the users consider the application easy to use. As the users are more equilibrated emotionally, their results are improved. This allowed the persons to be more confident on themselves. In the case of students, the neurofeedback can be studied for the better sport and artistic performances, including the case of the attention deficit hyperactivity disorder. Aptitudes for a subject can be determined based on the relevant generated brainwaves. The learning environment is an important factor during the analysis of the results. Teachers, professors, students and parents can collaborate and, based on the gathered data, new teaching methods can be adopted in the classroom and at home. The proposed solution can guide the students while studying, as well as the persons who wish to be more productive while solving their tasks.

*Keywords*—Neurofeedback; headset; electroencephalogram; brain performance

## I. INTRODUCTION

At the intersection point between neuroscience and information science, neuroinformatics is a research field that provides computational tools, mathematical models and interoperable databases to organized specific data. Neuroinformatics facilitates how the sub-disciplines of neuroscience share their data [1].

While there are many neuroinformatic projects and laboratories, some of their main goals are to research the organization of brain networks involved in human thoughts (University of Melbourne, Australia) [2], understanding the brain through neuroimaging (McGill University, Canada and The THOR Center of Neuroinformatics, Denmark) [3] or to discover how the brain works and further apply it on artificial systems (The Institute of Neuroinformatics, Switzerland) [4]. At the moment there are not many systems (such as SenseLab [5] or BrainML [6]) that provide useful data for further research.

The aim of the current paper is to introduce another system that collects brain performance data which can be later analysed by specialists. More specifically, the proposed system gathers the data from an Emotiv Insight electroencephalogram headset that reads the waves transmitted by the brain and quantifies parameters such as stress and focus.

The input data are interpreted and the system outputs some recommendations based on the results. In this way, the user can experience different situations and check the triggered results, based on which the personal behavior can be adapted for optimum results to improve the physical and mental wellbeing.

In the next section are presented the literature review regarding neuroscience, along with the proposed system of stress detection and reduction. Section 3 outlines the results after testing the system. The last section outlines the conclusions and future work.

## II. METHODOLOGY

The domain of neuroscience is further analyzed from the point of view of linguistics, arithmetic, artificial intelligence (AI), mindfulness based stress reduction, cortisol stress reactivity reduction and biofeedback. In addition, the proposed system is described.

### A. Linguistic and neuroscience

By creating a bridge between language and neuroscience, researchers aim to better understand the computational and encoding power of different areas of the brain [7]. The joint study has produced promising results that make the humans understand themselves more.

Several studies regarding Broca's area, a region widely known to be linked to speech reproduction, was declared that it specializes syntax [8]. In the study, the participants were asked to judge sentences depending on whether they belong to the 'syntactic' or 'semantic' category. Using the functional magnetic resonance imaging (fMRI) technology, an activation to the stimuli was found in the Broca area.

Further studies prove this claim to be oversimplified. Broca's area was seen to be active also when a participant was making lexical decisions or on studies regarding tonal languages [8]. This issue seems to come from two problems, namely the granularity mismatch which signify the mismatches between elemental concepts of language and neurobiology, leading to the prevention of formulation of theory biological grounded [9].

Another problem is given by the ontological incommensurable problem where the units of language computation and neurobiology computation are incommensurable [10].

### B. Cognitive neuroscience of arithmetic

Studies regarding the cognitive processes underlying the development of arithmetic skills have provided significant information regarding the human brain, especially the differences between children and adults [11]. This study aims to provide the foundation of the bridge between neuroscience and education. The focus is put on arithmetic skills, but, also, on other cognitive processes that influence it, namely calculation, retrieval, strategy use, decision making, working memory and attention.

Relevant research of adults' brains using fMRI have proved that acalculia, a difficulty in making even simple computation, appears in the case of people with deficits in the posterior parietal cortex, but also for those that suffered injuries in the prefrontal cortex [12]. Deficits in other regions of the brain proved to affect only one mathematical skill, such as subtraction, which suggests that arithmetical skills are spread through the brain.

Further study has proved that adults show a decreased activation to stimuli due to experience and exposure, and, for what is considered colloquially, 'simple math' such as one digit multiplication, they retrieve their answer from memory, in contrast to the children who show a great activation on areas linked to computation and other reconstructive procedures [13].

### C. Neuroscience inspired Artificial Intelligence

In the beginning, the fields of neuroscience, psychology and AI research were strongly dependent of each other, yet, in more recent years, the rapid progress of science reduces the communication and collaboration between them. Some researchers believe that is still not the time to give up on this connection since the biological brain might play a vital role in the development of artificial intelligence [14, 15].

The neuroscience provides a rich source of inspiration for algorithms and computational systems which are independent of the mathematical and logic-based ones used in the present [14]. Also, neuroscience provides the validation of what was discovered. To help the development of AI, the biological brain is a guide, while the researchers are interested in algorithms, architecture, functions and representations.

AI was revolutionized by deep and reinforced learning, but there are still gaps to fill, that might be by neuroscience, such as in the case of attention, when the brain does not apply the same optimization algorithms to every data as a neural network (NN). It should paid attention to certain details, like episodic memory where the intelligent behavior relies on more memory systems [16]. Also, the stimuli should be learnt and encoded.

The working memory maintains and manipulates information. Continual learning concerns new information, but it remembers the old one, unlike NN that after going to a more optimal stage, it forgets the last one. This is the greatest challenge of AI researchers. The bridge between neuroscience and AI is two-way. The intelligent system provides a much easier analysis of neuroimaging.

### D. Mindfulness based stress reduction

In recent years, 64 empirical studies were found that used mindfulness based stress reduction (MBSR) that employs an 8-week treatment with mindfulness meditation to alleviate sufferings such as anxiety and panic disorders [17]. The program is based on a procedure that develops enhanced awareness of the mental processes experiences. The research states that greater awareness will provide more veridical perception, reducing negativity and improving vitality.

Out of the 64 studies, only 20 were proved to be acceptable and relevant. Those studies cover participants with clinical problems due to pain, cancer, depression, but also participants with non-clinical ones. The results show that MBSR may help individuals to cope with their problems.

### E. Specific reduction in cortisol stress reactivity

While the MBSR methodology proved to be efficient, the studies are restricted and not recognized by everyone. More precisely, the cortisol told to be the most important marker is not affected by 8 weeks of therapy [18].

ReSource Project has done a research of a much longer period of time, 9 months, and with a significant number of individuals, 332 participants [18]. This project is composed of an analysis of three stages: presence, namely the present moment where attention is focused, affect which is characterized by gratitude, compassion, social motivation and difficult situations and perspectives which are characterized by taking self and other's perspectives.

During those months, participants took the Trier Social Test. Each group did hold the test during different stages, including before and after the 9 months. Various stress markers were used and, the most important marker, cortisol, had the secretion decreased with even 51%, especially in the second stage, indicating a clear reduction of the sociopsychological stress which contributes towards negative mental and physical health outcomes.

### F. Smartphone applications using biofeedback

The study done by Dillon et al. aims to bring a new stress-reduction method that not only is engaging but also appeals to the younger generation [19]. Biofeedback detects and presents physiological signals such as heart rate, the temperature of the skin, respiration, thus making the user aware of these symptoms and recognizing the situation.

The stress programs are, usually, done in peaceful environments which do not resemble the day-to-day life to solve this inconvenience, the researches came up with video games which contain a competitive element. From teaching the children to breathe, to helping soldiers cope with their post-traumatic stress disorder (PTSD), these games showed promising results.

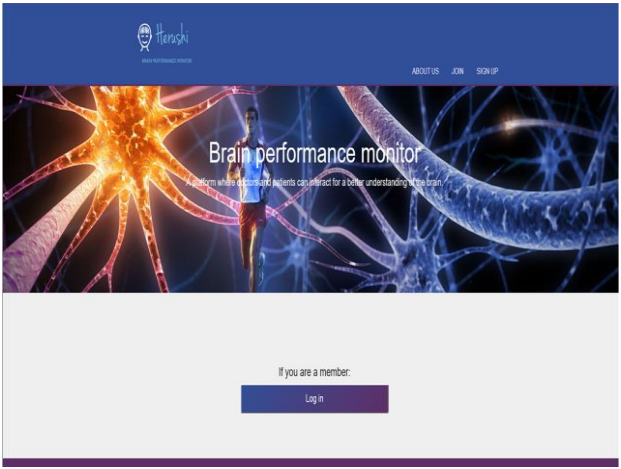

Fig. 1. *Application homepage*

With the help of mobile phones, stress programs have become more portable than ever. The participants took a trier social stress test before a 30 minutes game therapy when a Bluetooth connected biofeedback device gathers data and sends it for further analysis. After the therapy, the individuals showed significant stress reduction in a short period of time.

*G. The proposed system*

First of all, the paper's proposed system is meant to monitor the patient's brain over an average period of time of 3 days, on different moments of the day. This implies that the patient is not confined into a room or laboratory for several hours, but is at home, at work, at school, living their everyday life.

The stressful situations are not mocked by a trier social stress test, thus they are what the patient is dealing with on a daily basis. The recommendations provided by the system are based on a try and error principle. It was realized that the same method may function only on a number of people, so patients are encouraged to try more approaches to find the one that fits them the best.

## III. RESULTS

The patient wears the specially designed headset, the Emotiv Insight electroencephalogram headset, whose sensors read and quantify the electromagnetic waves send by the brain. The headset sends the interpreted data to a mobile application that displays the results.

The patient inserts the triggered neurofeedback data in the application, namely activity, mood, focus, stress, relaxation, engagement, excitement and interest. The application immediately stores the data and displays the new results on the profile page.

In order to access the platform, the user must be logged in, as displayed in Fig. 1. As soon as the web application is opened, the home page redirects the user to a login page, if it is not the first time when the application is used or a sign in page for the persons without an account.

The user is redirected to the personal profile page where, the statistics are displayed, along with a list of recommendations and the option to add data. The list of recommendations can be further exploited by the user in the process of learning, as well as solving tasks at work.

The users can add activities related to relaxation, listening to music, watching a movie, or studying. These activities can be to practice a hobby, do more sport, walk in the forest, do manual work like sewing and chores like ironing. The learning environment is a key factor while analyzing the brainwaves based on the weighted average of the recorded results.

When adding data, as in Fig. 2, the patients must be careful to also insert the neurofeedback values for activity, mood, focus, stress, relaxation, engagement, excitement and interest, because they also affect the triggered results. The values are in the range from 0 to 100.

The users of the platform have completed a short feedback and it was found that it is regarded as acceptable and quite easy to navigate, but the main problem is that the data must be inserted manually.

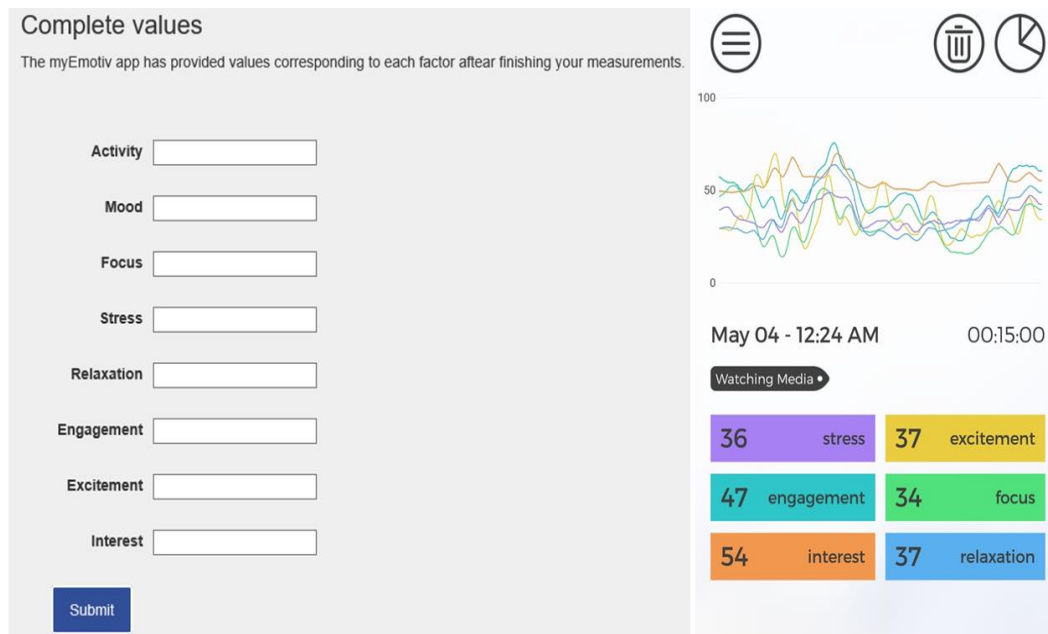

Fig. 2. *Data analysis*

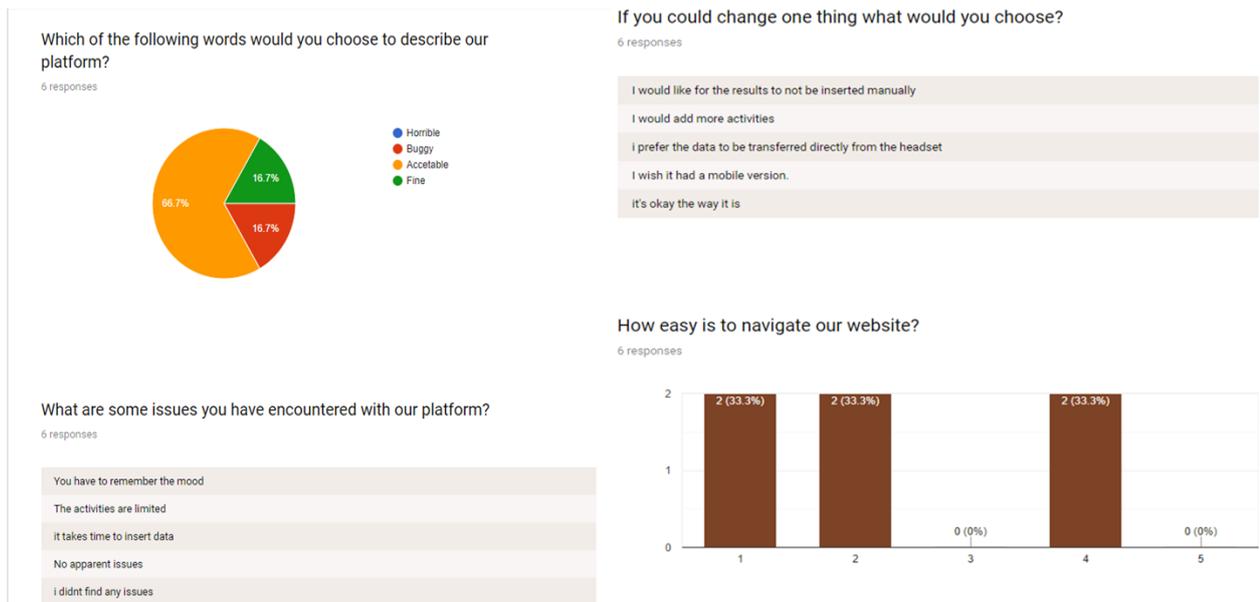

Fig. 3. *Feedback from users*

The results of the feedback are displayed in Fig. 3. As the users continued to test their brainwaves, while equilibrating themselves by performing relaxing activities, their scores improved and they believed more in themselves. Wavelengths can be an important mean for determining the student aptitudes for different subjects.

IV. CONCLUSIONS

Neurofeedback can be successfully used by people who work in multiple domains, because behavioral problems like deficit of attention and engagement are much more common these days. The day to day activities are diverse and closely connected to the triggered measurements performed with the use of an electroencephalogram headset.

The article aims to present a conception created to facilitate the analysis of the brain's performance, thus aiding teachers and professors in helping their students in a more effective manner to combat stress in their everyday life.

The presented application can be an effective education tool that can be used by students, while being at school or university. The teachers, professors, students and parents will be closely connected and a personalized studying plan can be created. During the monitoring evolution process, further adjustments of the teaching activities can be performed.

The future improvements aim to increase the database of activities, improve security and optimize the algorithms. Also, more persons will be involved in order to determine a teaching strategy which will suit most students who attend classes.

REFERENCES

[1] S. Adee, "Reverse engineering the brain," IEEE Spectrum, vol. 45(6), pp. 51-53, 2008.

[2] A. Kottaram, L. Johnston, L. Cocchi, E. Ganella, I. Everall, C. Pantelis, R. Kotagiri and A. Zalesky, "Brain network dynamics in schizophrenia: Reduced dynamism of the default mode network," Human Brain Mapping, vol. 40(7), 2019.

[3] MR Core Research Facility, "The Magnetic Resonance Imagining Institute for Biomedical Research," http://www.mrc.wayne.edu/mrlinks, accessed 20.02.2020.

[4] B. Wrobel, A. Abdelmotaleb and M. Joachimczak, "Evolving Networks Processing Signals with a Mixed Paradigm, Inspired by Gene Regulatory Networks and Spiking Neurons, Bio-Inspired Models of Network, Information, and Computing Systems," 7th International ICST Conference, 2014.

[5] SenseLab, "A laboratory for thought in motion," https://senselab.ca, accessed 20.02.2020.

[6] BrainML, "A standard XML metaformat for exchaning neuroscience data," http://www.brainml.org, accessed 21.02.2020.

[7] D. Poeppel and D. Embick, Defining the relation between linguistics and neuroscience, Lawrence Erlbaum Associates, 2017, pp. 103-118.

[8] C. Cantalupo and W. Hopkins, Asymmetric Broca's area in great apes, Nature, 2001.

[9] B. Wong, B. Yin and B. O'Brien, "Neurolinguistics: Structure, Function, and Connectivity in the Bilingual Brain," BioMed Research International, 2016.

[10] M. Grimaldi, "Toward a neural theory of language: Old issues and new perspectives," Journal of Neurolinguistics, vol. 25(5), pp. 304-327, 2012.

[11] V. Menon, "Developmental cognitive neuroscience of arithmetic: implications for learning and education," ZDM: The International Journal on Mathematics Education, vol. 42(6), pp. 515-525, 2010.

[12] A. Ardila and M. Rosselli, Acalculia and Dyscalculia, Neuropsychology Review, vol. 12(4), pp. 179-231, 2003.

[13] H. Aini Ismafairus Abd, Y. A. Nazlim, S. M. S. Zamratol-Mai and M. Mazlyfarina, "Brain Activation during Addition and Subtraction Tasks In-Noise and In-Quiet," The Malaysian Journal of Medical Sciences, vol. 18(2), pp. 3-15, 2011.

[14] D. Hassabis, D. Kumaran, C. Summerfield and M. Botvinick, "Neuroscience-Inspired Artificial Intelligence," Neuron, vol. 95(2), pp. 245-258, 2017.

[15] P. Shapshak, "Artificial Intelligence and Brain," Bioinformation, vol. 14(1), pp. 38-41, 2018.

[16] M. Moscovitch, R. Cabeza, G. Winocur and L. Nadel, "Episodic Memory and Beyond: The Hippocampus and Neocortex in Transformation," Annual Review of Psychology, vol. 67, pp. 105-134, 2016.

[17] P. Grossman, L. Niemann, S. Schmidt and H. Walach, "Mindfulness-based stress reduction and health benefits," Journal of Psychosomatic Research, vol. 57(1), pp. 35-43, 2004.

[18] V. Engert, B. Kok, I. Papassotiriou, G. Chrousos and T. Singer, "Specific reduction in cortisol stress reactivity after social but not attention-based mental training," Science Advances, vol. 3(10), 2017.

[19] A. Dillon, M. Kelly, I. Robertson and D. Robertson, "Smartphone Applications Utilizing Biofeedback Can Aid Stress Reduction," Frontiers in Psychology, vol. 7, 2016.